\documentclass[10pt]{article}
\usepackage{amssymb}
\usepackage{amsmath}
\usepackage{amsthm}
\usepackage{latexsym}
\usepackage{mathrsfs}
\usepackage{bm}
\usepackage{eufrak}
\usepackage{tikz}
\usetikzlibrary{calc,through,backgrounds,decorations}
\usepgflibrary{decorations.pathmorphing} 
\usepgflibrary{decorations.pathreplacing}
\usetikzlibrary{arrows}

\theoremstyle{plain}

\newtheorem{lemma}{Lemma}
\newtheorem{theorem}{Theorem}

\newtheorem{corollary}{Corollary}

\font\tenscr=rsfs10 scaled1100
\font\sevenscr=rsfs7 
\font\fivescr=rsfs5 
\skewchar\tenscr='177
\skewchar\sevenscr='177
\skewchar\fivescr='177
\newfam\scrfam
\textfont\scrfam=\tenscr
\scriptfont\scrfam=\sevenscr
\scriptscriptfont\scrfam=\fivescr

\def\Scri{{\fam\scrfam I}}

\newcommand{\tensor}[3]{_{#1\phantom{#2}#3}^{\phantom{#1}#2}}

\newcommand{\cg}{\mathbf{g}}

\newcommand{\sect}{\varepsilon}

\newcommand{\cv}{\nu} 

\newcommand{\D}{\nabla}

\newcommand{\Schouten}{L}
\newcommand{\cSchouten}{\mathbf{L}}

\newcommand{\BoxOp}{\square}

\newcommand{\gtilde}{\tilde{g}}

\newcommand{\Schoutentilde}{\tilde{L}}

\newcommand{\gup}{\hat{g}}
\newcommand{\cfup}{\hat{\theta}}
\newcommand{\Omegaup}{\hat{\Omega}}

\newcommand{\Dup}{\hat{\nabla}}

\newcommand{\Schoutenup}{\hat{L}}
\newcommand{\cSchoutenup}{\mathbf{\hat{L}}}
\newcommand{\Tup}{\hat{T}}
\newcommand{\Mup}{\hat{M}}
\newcommand{\Gup}{\hat{G}}

\newcommand{\BoxOpup}{\hat{\square}}

\newcommand{\gdown}{\check{g}}
\newcommand{\cfdown}{\check{\theta}}
\newcommand{\Omegadown}{\check{\Omega}}

\newcommand{\Ddown}{\check{\nabla}}

\newcommand{\Tdown}{\check{T}}
\newcommand{\Mdown}{\check{M}}
\newcommand{\Gdown}{\check{G}}

\newcounter{mnote}

\newcommand{\cs}{\hat{\sigma}} 
\newcommand{\scalarfield}{\hat{\varphi}} 
\newcommand{\dualscalarfield}{\check{\varphi}} 
\newcommand{\dualcs}{\check{\sigma}} 
\newcommand{\Cosmolconstup}{\hat{\Lambda}} 
\newcommand{\cosmolconstup}{\hat{\lambda}} 
\newcommand{\CouplingConst}{\alpha} 
\newcommand{\cosmolconstdown}{\check{\lambda}} 
\newcommand{\Phiup}{\hat{\Phi}}
\newcommand{\phiup}{\hat{\phi}}
\newcommand{\Phidown}{\check{\Phi}}
\newcommand{\phidown}{\check{\phi}}
\newcommand{\EMtensordensity}{\mathbb{T}}
\newcommand{\EMoperator}{\mathbb{D}}

\begin{document}

\title{\textbf{Conformal scalar fields, isotropic singularities and conformal cyclic cosmologies}}

\author{
{\Large Christian L\"ubbe} \thanks{E-mail address:
 {\tt christian.luebbe@gmail.com, \tt c.luebbe@ucl.ac.uk}}\\ ~\\
{ Department of Mathematics, University College London, London, UK} \\
{ Graduate School of Mathematical Sciences, University of Tokyo, Tokyo, Japan}
}

\maketitle

\begin{abstract}
We analyse spacetimes with a conformal scalar field source, a cosmological constant and a quartic self-interaction term for the scalar field. We also consider additional matter contents in the form of Maxwell and Yang-Mills fields or radiation fluids. Existence theorems for weakly asymptotically flat spacetimes are given. We give a generalisation of Bekenstein's result [Ann. Phys. 82, 535 (1974)] and use it to derive existence theorems for spacetimes that contain an isotropic singularity.
The results are combined to suggest a mathematical setup for Penrose's CCC scenario using a conformal scalar field cosmology.
\end{abstract}

\section{Introduction}
Penrose's work on conformal geometry and general relativity \cite{Penrose-LesHouchesLectures}, \cite{Penrose ZRM} (see also \cite{Penrose-GoldenOldie}
) has become an established approach to study the asymptotic structure of spacetimes. The central idea is to study a physical spacetime $(\tilde{M}, \gtilde)$ and its asymptotic structure in terms of a conformally related spacetime $(M,g)$, with $g = \theta^2 \gtilde $. The Einstein field equations are not satisfied in $(M,g)$ and hence $(M,g)$ is typically referred to as the unphysical spacetime. The conformal boundary of $(\tilde{M}, \gtilde)$ is given by the set where the conformal factor $\theta$ vanishes. The conformal approach allows one to study $(\tilde{M}, \gtilde)$ in terms of the properties of fields at the conformal boundary.

The treatment of isotropic singularities 
\cite{Tod IS definition}, also known as conformally compactifiable singularities \cite{AngTodperfectfluid} or conformal gauge singularities \cite{LueTod conf gauge sing}, is similar in spirit to the conformal approach for the asymptotic structure. Again the physical spacetimes $(\tilde{M}, \gtilde)$ is conformally embedded into a larger unphysical spacetime $(M,g)$, with $g = \theta^2 \gtilde $. However this time the conformal factor $\theta$ diverges as one approaches the singularity. 
In \cite{AngTodperfectfluid}, \cite{AngTodVlasov}, \cite{Tod-Torino} isotropic singularities with different matter models were studied to investigate the Weyl tensor hypothesis proposed by Penrose \cite{Penrose PWTH}, \cite{Penrose Majid}, \cite{Penrosebook}. 
Penrose argued that at the big bang the gavitational entropy should have been very low and that the Weyl tensor at the big bang should have vanished (strong Weyl curvature hypothesis) or at least been non-singular (weak Weyl curvature hypothesis). In the case of a perfect fluid with $p=(\gamma-1)\rho$, Anguige and Tod showed in \cite{AngTodperfectfluid} that an initially vanishing Weyl tensor implies that the spacetime is globally conformally flat. In the case of massless Vlasov the same authors \cite{AngTodVlasov}  showed that there exist spacetimes for which the Weyl tensor vanishes at the isotropic singularity but which are not conformally flat. In \cite{AngTodperfectfluid}, \cite{AngTodVlasov} a vanishing cosmological constant was used, while the case with a de Sitter-like cosmological constant was analysed for spatially homogeneous spacetimes in \cite{Tod-IS deSitter}.

In \cite{Penrose Majid}, \cite{Penrosebook} Penrose outlined details of his recent proposal of conformal cyclic cosmologies (in the following CCC). At the centre of the CCC-proposal lies the idea that spacetimes (termed aeons) with a de Sitter-like cosmological constant form a successive chain. Two consecutive aeons are joint in a \emph{bridging spacetime} \cite{Tod2013CCC} by identifying the future null infinity of one aeon with the isotropic singularity describing the big bang of the next aeon. One aspect that has been highlighted in \cite{Tod ERE} is that the matching of $\Scri$ imposes a vanishing Weyl tensor, while the Weyl tensor may be non-zero at the isotropic singularity. Thus the setting in the future aeon appears to require specific fine tuning. 
The results of \cite{AngTodperfectfluid}, \cite{AngTodVlasov} highlight that for a chosen matter model we need to check whether there exists a sufficiently large family of solutions satisfying the fine-tuning or whether one is automatically reduced to the conformally flat case. The existence of explicit pairs of physical spacetimes satisfying the CCC-proposal which are not conformally flat has been recently shown in \cite{Tod2013CCC}. \footnote{New conformally flat solutions were found in \cite{Newman2013CCC}.} 

Observe that, despite the use of an unphysical \textit{'bridging metric'} to formulate the concepts of conformal boundary and isotropic singularity, the CCC-scenario requires two physical spacetimes that are conformally related.
A similar conformal relationship between two physical spacetimes was already observed by 
Bekenstein in \cite{Bekenstein1974}. He showed that a spacetime $(\tilde{M}, \gtilde)$ with an ordinary scalar field, an electromagnetic field and a radiation fluid is conformally related to a spacetime $(\hat{M}, \hat{g})$ containing a conformal scalar field, an electromagnetic field and a radiation fluid. It was also shown that if the spacetime $(\tilde{M}, \gtilde)$ contains only an ordinary scalar field then it is conformally related to two spacetimes $(\Mup,\gup)$ and $(\Mdown,\gdown)$ each containing a conformal scalar field. Unlike $(M,g)$, the manifolds $(\tilde{M}, \gtilde)$, $(\Mup,\gup)$ and $(\Mdown,\gdown)$ are all physical solutions in their own right, using different matter models. 
Bekenstein's spacetimes have the advantage that there is a clear mathematical procedure that generates a new physical spacetime from a given one. Since the work in \cite{Bekenstein1974} is only concerned with $\Lambda=0$ we will extend the results to include a non-zero cosmological constant. Moreover, we will investigate how such a result could provide new ideas for generating the new aeon from the previous one.

\subsection{Main results}
 Our results can be summarised as follows
 
 \medskip

\noindent \textbf{Main Theorem:} Let $\Sigma$ be a compact spacelike hypersurface. Suppose on $\Sigma$ we are given initial data at null infinity for the CEFE with a de Sitter-like cosmological constant whose matter model is a conformal scalar field $\phiup$ with quartic self-interaction term minimally coupled to Einstein-Maxwell-Yang-Mills and radiation fluids. Then the following hold:
\begin{enumerate}
\item  There exists a weakly asymptotically flat spacetime $(\Mup, \gup) $ with a conformal scalar field minimally coupled to Einstein-Maxwell-Yang-Mills and radiation fluids.
\item  There exists a second solution $(\Mdown, \gdown) $ with the same matter models for which the values of the cosmological constant and the coefficient of the quartic self-interaction term are interchanged. If the unphysical scalar field $\Phiup=\phiup / \theta$ vanishes exactly on $\Sigma$ then $(\Mdown, \gdown) $ has an isotropic singularity at $\Sigma$ that satisfies the strong Weyl curvature hypothesis. 
\item $(\Mup, \gup) $ and $(\Mdown, \gdown) $ can be interpreted as consecutive aeons of Penrose's conformal cyclic cosmology.
\end{enumerate} 

For the detailed assumptions, precise formulations and technical details
the reader is referred to the main text.

\subsection{Outline}

We start by setting up the necessary geometry and notation in Section \ref{Conformal geometry}. In particular, we extend Bekenstein's results \cite{Bekenstein1974} on the duality of  conformal scalar field spacetimes to include a cosmological constant and a quartic self-interaction term. 
In Section \ref{WAFST} we discuss the existence of weakly asymptotically flat spacetimes containting a conformal scalar field coupled with Yang-Mills fields and radiation fluids. Using the duality between conformal scalar field spacetimes we derive existence results for spacetimes contain an isotropic singularity and whose matter is given by a conformal scalar field, Einstein-Maxwell-Yang-Mills and radiation fluids.
In Section \ref{CCC-proposal} we show that these spacetimes can be joined to form consecutive aeons of the CCC-scenario and discuss some of the questions that arise in this context. We conclude with a brief discussion and some comments.

\section{The conformal geometry of the conformal scalar field}
\label{Conformal geometry}

The conventions and notations in this article are those of \cite{LueTod conf gauge sing}. We briefly summarise the most important ones. 
Throughout we work in $n=4$ dimensions and the metrics have signature $(+---)$. 
The bundles of scalar conformal densities of conformal weight $w$ is denoted $\sect[w] $. They can be used to define more general conformal densities by $\sect_{{\cal{A}}}[w] := \sect_{{\cal{A}}} \otimes \sect[w] $, where $\cal{A}$ is some general bundle index. $\sect[1]$ is the bundle of conformal scales. 
Let $\cs\in \sect[1]$ denote the physical conformal scale. Then the physical metric is given by $\gup_{ij}=\cs^{-2} \cg_{ij} $, where $\cg_{ij} \in \sect_{ij}[2]$ is the conformal metric. The associated Levi-Civita connection is denoted by $\Dup$ and satisfies $\Dup_i \cs=0$. We will denote the unphysical metric by $g$ and the associated conformal scale and Levi-Civita connection by $\cv$ and $\D$. Our curvature conventions are $(\D_i \D_j - \D_j \D_i) v^k = R\tensor{ij}{k}{l}  v^l$ and $R_{jl}=R\tensor{kj}{k}{l} $. The Schouten tensor for a Levi-Civita connection is given by $L_{ij} = \tfrac{1}{2} (R_{ij} - \tfrac{1}{6} R g_{ij}) $.
When working with trace-free matter models we will use the rescaled energy-momentum tensor $T_{ij} = \cfup^{-2}\Tup_{ij} $, which also satisfies $\D^i T_{ij} $ \cite{FriEMYM}. The associated conformal density is given by $\EMtensordensity_{ij} = \cs^{-2} \Tup_{ij} \in \sect_{ij}[-2] $.

\subsection{The physical setting}

Let $\scalarfield \in \sect[-1]$ be the conformal density representing the conformally scalar field.  Observe that $\phiup := \cs \scalarfield $ is the representation of $\scalarfield$ in the conformal scale $\cs$ and thus the value of the conformal scalar field in the physical spacetime. 

The energy momentum tensor for a conformal scalar field \cite{CallanColemanJackiw}, \cite{Parker1973} can be given by
\begin{equation}
\label{EMtensor conf inv scalar field}
\epsilon \Tup_{ij} =  4\Dup_i \phiup \Dup_j \phiup - \gup_{ij}\Dup^k \phiup \Dup_k \phiup - 2\phiup \Dup_i \Dup_j \phiup
+ 2\phiup^2 \Schoutenup_{ij} - 2 \CouplingConst \phiup^4 \gup_{ij}
\end{equation}
where $\epsilon = \pm 1$. We refer to the case $\epsilon=1$ as the (attractive) conformal scalar field and to $\epsilon=-1$ as the repulsive conformal scalar field. We note that conformal scalar field can violate the null energy condition and hence the other energy conditions. The last term in \eqref{EMtensor conf inv scalar field} will be referred to as the quartic self-interaction term.
Imposing that $\phiup$ satisfies the inhomogenous conformally invariant wave equation $(\Schoutenup = \Schoutenup_{ij}\gup^{ij})$
\begin{equation}
\label{physical conf wave equation}
\BoxOpup \phiup - \Schoutenup \phiup = - 4 \CouplingConst \phiup^3
\end{equation}
implies that $\Tup_{ij}$ is trace-free and divergence free. Note that the coefficients of the quartic interaction term in \eqref{EMtensor conf inv scalar field} and of the cubic in \eqref{physical conf wave equation} have to coincide.
The related Einstein field equation with a cosmological constant $\Cosmolconstup$ is given by
\begin{eqnarray}
\label{EFE}
\Gup_{ij} + \Cosmolconstup \gup_{ij}= \Tup_{ij}.
\end{eqnarray}
We observe that $\Gup_{ij} = 2(\Schoutentilde_{ij} - \Schoutenup \gup_{ij}) $ and set $\Cosmolconstup = 6 \cosmolconstup$ for later convenience. As $\hat{T}_{ij}$ is tracefree we have $\Schoutenup = 4 \lambda $ so that \eqref{EFE} can be written as
\begin{equation}
\label{rewritten EFE}
2\Schoutenup_{ij} -   2\cosmolconstup \gup_{ij}= \Tup_{ij}
\end{equation}
and \eqref{physical conf wave equation} takes the form 
$$
\BoxOpup \phiup - 4\cosmolconstup \phiup = - 4 \CouplingConst \phiup^3.
$$
This can be interpreted in terms of a potential $V=V(\phiup)$ and written as 
$$ 
\BoxOpup \phiup + \frac{\partial V}{\partial \phiup} =0 \quad\quad \mathrm{with} \quad\quad V-V_0 =  \CouplingConst   \left(   \phiup^4 - 2 \frac{\cosmolconstup}{\CouplingConst}  \phiup^2  \right) =  \CouplingConst   \left(   \phiup^2 -  \frac{\cosmolconstup}{\CouplingConst} \right)^2 -\frac{\cosmolconstup^2}{\CouplingConst}.
$$
If the spacetime is de Sitter-like, i.e. $\cosmolconstup<0$, then for $\CouplingConst>0$ the potential has a global minimum at $\phiup=0$
and for $\CouplingConst<0$ we have an inverted mexican hat potential.
In the case $\CouplingConst=0$ the potential reduces to $V = -2 \cosmolconstup \phiup^2 $.

Isolating $\Schoutenup_{ij}$ from \eqref{rewritten EFE}, substituting into \eqref{EMtensor conf inv scalar field} and isolating $\Tup_{ij}$ allows one to find the commonly used expression for the physical energy-momentum tensor in terms of $\phiup, \cosmolconstup, \CouplingConst$ only
\begin{equation}
\label{EMtensor scalar field only}
\Tup_{ij} = ( \epsilon - \phiup^2)^{-1}\left[ 4\Dup_i \phiup \Dup_j \phiup - \gup_{ij}\Dup^k \phiup \Dup_k \phiup - 2\phiup \Dup_i \Dup_j \phiup
+ 2 \cosmolconstup\phiup^2 \gup_{ij} - 2 \CouplingConst \phiup^4 \gup_{ij} \right] .
\end{equation}
For $\epsilon=1$ \eqref{EMtensor scalar field only} holds as long as $\phiup \ne 1 $. When $\phiup \in (-1,1) $ the factor at the front is positive, whereas when $\vert \phiup \vert > 1 $ the factor is negative.
For $\epsilon=-1$ \eqref{EMtensor scalar field only} holds everywhere and the factor is always negative.

There exists a trivial solution $\phiup=0$, which describes  vacuum. For a constant scalar field $\phiup=\phiup_0\ne 0$ equation \eqref{physical conf wave equation} gives $\phiup_0^2 = \frac{\cosmolconstup}{\CouplingConst} $ which in turn implies that $T_{ij}=0$ and thus also describes vacuum. Since we are not interested in vacuum solutions we will throughout assume that $\Dup_i \phiup \ne 0$ or $\Dup_i \Dup_j  \phiup \ne 0$ when $\phiup=0$

\subsection{The conformal setting and some conformal properties}
We say a connection is compatible with $[\cg]$ if $\D_i\cg_{jk}=0$ and $\D$ is torsion-free. For a metric $\gup\in[\cg] $ the Levi-Civita connection $\Dup=\D^{(\gup)} $ is compatible with $[\cg]$, as are all the general Weyl connections $\D$. 

In the following we work in the language of conformal densities as this saves us carrying explicit conformal factors through all the calculations. For $\scalarfield \in \sect[-1]$ the conformally invariant wave equation has the form
\begin{equation}
\label{conf inv wave equ}
\BoxOp \scalarfield - \cSchouten \scalarfield= - 4 \CouplingConst \scalarfield^3,
\end{equation}
where $\D$ is any connection compatible with $[\cg]$, $\BoxOp=\cg^{ij}\D_i\D_j $ and $\cSchouten=\Schouten_{ij}\cg^{ij}$ is the conformal trace of the Schouten tensor of $\D$.

Given a conformal density $\scalarfield \in \sect[-1]$, a connection $\D$ compatible with $[\cg]$ and its Schouten tensor $L_{ij}$ as well as a parameter $\CouplingConst $ we define the following conformal density
\begin{equation}
\label{EMtensordensity}
\EMoperator_{ij}[\scalarfield, \D, \CouplingConst] := 4 \D_i \scalarfield \D_j \scalarfield - \cg_{ij}\D^k \scalarfield \D_k \scalarfield - 2 \scalarfield \D_i \D_j \scalarfield + 2 \scalarfield^2 \Schouten_{ij} -2 \CouplingConst \scalarfield^4 \cg_{ij} \in\sect_{ij}[-2]
\end{equation}
We observe the following properties of $\EMoperator_{ij}[\scalarfield, \D, \CouplingConst] $
\begin{enumerate}
\item $\EMoperator_{ij}[C\scalarfield, \D, \CouplingConst] =C^2 \EMoperator_{ij}[ \scalarfield, \D, \CouplingConst/C^2] $

\item If $\D$ and $\Dup$ are both compatible with $[\cg]$ then
$$
\EMoperator_{ij}[\scalarfield, \D, \CouplingConst] = \EMoperator_{ij}[\scalarfield, \Dup, \CouplingConst].
$$
Thus we can simply write the tensor density as $\EMoperator_{ij}[\scalarfield, \CouplingConst] $.

\item Taking the trace we have $ \cg^{ij} \EMoperator_{ij}[\scalarfield, \CouplingConst] = -2\scalarfield \left( \square \scalarfield - \cSchouten \scalarfield + 4\CouplingConst \scalarfield^3 \right)$. 

Thus if $\scalarfield$ satisfies \eqref{conf inv wave equ} then 
$\EMoperator_{ij}[\scalarfield, \CouplingConst]$ is trace-free. Moreover, \eqref{conf inv wave equ} implies that $\EMoperator_{ij}[\scalarfield, \CouplingConst]$ is divergence-free.

\item If $\cs$ denotes the physical scale, $\phiup = \cs \scalarfield $ and $\Dup$ satisfies $\Dup_i \cs=0$ then we have \footnote{Setting $C=\pm i$ in 1.) switches between the cases $\epsilon=1$ and $\epsilon=-1$. However, it also transforms a real scalar field into a purely imaginary one.}
\begin{equation}
\label{EFE RHS}
\cs^2 \EMoperator_{ij}[\scalarfield, \CouplingConst] = \epsilon \Tup_{ij}.
\end{equation}

\item Let $\dualscalarfield := \cs^{-1}$ then $\dualscalarfield$ satisfies
\begin{equation}
\label{EFE LHS}
\EMoperator_{ij}[\dualscalarfield, \cosmolconstup]  =2 \dualscalarfield^2 \Schoutenup_{ij} -2 \cosmolconstup \dualscalarfield^4 \cg_{ij}
=2\dualscalarfield^2 ( \Schoutenup_{ij} - \cosmolconstup \gup_{ij})
\end{equation}
and 
$$ 
\square \dualscalarfield - \cSchouten \dualscalarfield =
\hat{\square} \dualscalarfield - \cSchoutenup \dualscalarfield = 
- 4\cosmolconstup \dualscalarfield^3
$$
\end{enumerate}
Combining \eqref{rewritten EFE}, \eqref{EFE RHS}, \eqref{EFE LHS} we find
$$
\cs^2 \EMoperator_{ij}[\dualscalarfield, \cosmolconstup] = 2 ( \Schoutenup_{ij} - \cosmolconstup \gup_{ij}) = \Tup_{ij} = \epsilon \cs^2 \EMoperator_{ij}[\scalarfield, \CouplingConst] 
$$
Thus the Einstein field equation can be recast into the conformally invariant equation \footnote{This is similar to equations (17) and (22) in \cite{Parker1973}}
\begin{equation}
\label{conformal EFE}
\EMoperator_{ij}[\dualscalarfield, \cosmolconstup] = \epsilon\, \EMoperator_{ij}[\scalarfield, \CouplingConst].
\end{equation}
The important fact to observe is that in \eqref{conformal EFE} we can interpret either side as the Einstein tensor and the other side as the energy-momentum tensor. The two corresponding physical metrics are given by $\gup_{ij}=\cs^{-2}\cg_{ij}$ and $\gdown_{ij}=\dualcs^{-2}\cg_{ij}$. 
Since $\dualscalarfield = \cs^{-1} $ and $\scalarfield = \dualcs^{-1} $ we have 
$\phiup=\scalarfield \cs = \dualcs^{-1}\cs $ and $\phidown=\dualscalarfield \dualcs = \cs^{-1}\dualcs = \phiup^{-1}$. Hence the physical metrics are related by
\begin{equation}
\gdown_{ij}  =\Omegaup^2 \gup_{ij} \,\,\mathrm{with}\,\, \Omegaup = \phiup 
\quad \textmd{and}\quad
\gup_{ij} =\Omegadown^2 \gdown_{ij} \,\,\mathrm{with}\,\,  \Omegadown = \phidown.
\end{equation}

The existence of two dual solutions to the conformal scalar Einstein field equations was already observed in Theorem 2 of \cite{Bekenstein1974}, where $\cosmolconstup=0=\CouplingConst$ was assumed. The above observations lead to a straight forward generalisation of Bekenstein's result and can be summarised as follows
\begin{theorem}
Let $(\Mup, \gup) $ denote a solution of the Einstein field equations with cosmological constant $\Cosmolconstup=6\cosmolconstup $ and conformal scalar field $\phiup $ satisfying \eqref{EMtensor conf inv scalar field} with coefficient $\CouplingConst$. 
Then there exist a dual solution $(\Mdown, \gdown)  $ with scalar field $\phidown=\phiup^{-1}$  and $\gdown_{ij}  =\phiup^2 \gup_{ij} $. 
Moreover, the role of the cosmological constant $\cosmolconstup  $ and the coefficient of the quartic self interaction term $\CouplingConst $ have been swapped.
\end{theorem}

\subsubsection{An example illustrating the duality}

In Appendix D of \cite{BST} the authors gave conformal scalar field spacetimes obtained from static vacuum spacetimes. In these examples $\Cosmolconstup=0$ and $\CouplingConst=0$. Equ. (D74) of \cite{BST} gives
\begin{eqnarray}
\label{example metric}
\mathrm{d}s^2 &=& \frac{1}{4}\left(W^\beta\pm W^{-\beta} \right)^2 \left[ W^{2\alpha} \mathrm{d}t^2- W^{-2\alpha}h_{ij} \mathrm{d}x_i \mathrm{d}x_j    \right]\\
\label{example conf scalar field}
\phi &=& C\,\, \frac{1 \mp W^{2\beta}}{1 \pm W^{2\beta}} = - C\,\, \frac{W^\beta\mp W^{-\beta} }{W^\beta\pm W^{-\beta}} .
\end{eqnarray}
The top sign gives a solution of type A, denoted $(\gup, \phiup)$ here, and the bottom sign a solution of type B, denoted $(\gdown, \phidown)$. It is straight forward to check that $\gdown=\phiup^2 \gup$ and $\phidown=\phiup^{-1}$ so that solutions of types A and B are dual to each other. Moreover, when $W=e^U=1$, i.e. $U=0$ we have $\phiup=0$ and $\phidown=\infty$. 

A spherically symmetric example is given by equ. (D74) of \cite{BST}
\begin{eqnarray}
\label{eRN metric}
\gup &=&  \left( 1-\frac{m}{\bar{r}}\right )^2 \mathrm{d}t^2- \left( 1-\frac{m}{\bar{r}}\right )^{-2} \mathrm{d}\bar{r}^2 - \bar{r}^2 \mathrm{d}\sigma_2   \\
\label{eRN conf scalar field}
\phiup &=& C\,\, \frac{m}{\bar{r}-m}
\end{eqnarray}
which is the metric of extremal Reissner-Nordstr\"om black hole.
Defining the isotropic radial coordinate $R=\bar{r}-m$ we get $\phiup=C\frac{m}{R} $. Clearly $\phiup=0$ at $R=\infty$ and  $\phiup=\infty$ at $R=0$. Observe that the conformal rescaling $\gdown = \phiup^2 \gup $ recovers the  discrete conformal isometry of the extremal Reissner-Nordstr\"om spacetime. This conformal isometry $i: \gup \to \gdown $ maps the spacetime onto itself by $i:R\to \rho=\frac{m^2}{R}$. In particular, the horizon $R=0$ is mapped to null infinity $R=\infty$ and vice versa --- see \cite{BizFri eRN}, \cite{LueVal eRN} for further details. So in a sense, the metric \eqref{eRN metric} is its own dual. As highlighted in \cite{BST} the geometry of the metric \eqref{eRN metric} is regular at the horizon $R=0$ despite the blow-up of $\phiup$. This highlights that the blow-up of $\phiup$ need not imply a singular geometry.

\subsection{Conformal scalar fields coupled to other matter models}

Let us consider the case where the conformal scalar field is coupled to other trace-free matter models, like Einstein-Maxwell-Yang-Mills, radiation fluids, null fluids or massless Vlasov. Throughout this article we will assume that this coupling to other matter is minimal, i.e. the energy momentum tensor for each individual matter component is divergence-free. 

We will be particularly interested in Einstein-Maxwell-Yang-Mills and radiation fluids as we will want to use the conformal Einstein field equations (CEFE) for both models, see \cite{FriEMYM} and \cite{LueVal radiation fluid}, to prove the existence of asymptotically flat solutions. We remark that our results generalise to combinations of multiple Yang-Mills fields, radiation fluids and conformal scalar fields, however in the interest of clarity we will only discuss the case of at most one component of each matter type. We would also like to highlight that the results concerning dualities of spacetimes generalise to more general trace-free models, like null fluids or massless Vlasov. However, a central part of our work will rely on the CEFE and the related existence and uniqueness results. At this moment in time there are no CEFE available for these matter models that allow us to formulate an IVP at conformal infinity and evolve the conformal spacetime, which is the main reason why models like null fluids or massless Vlasov have been excluded from our analysis.

Recall that the physical and unphysical energy momentum tensors $\hat{T}_{ij} $ and $T_{ij} $ are given by
\begin{eqnarray}
\Tup_{ij}^{[EMYM]} &=& F^\alpha\,_{il} F_\alpha\,\tensor{}{l}{j} - \frac{1}{4}g_{ij} F^\alpha\,_{kl} F_\alpha\,^{kl} = \cfup^{2} T_{ij}^{[EMYM]}\\ 
\Tup_{ij}^{[rad]} &=& \frac{4}{3} \hat{\rho} \hat{u}_i \hat{u}_j - \frac{1}{3} \hat{\rho} \gup_{ij} = \cfup^{2} T_{ij}^{[rad]} .
\end{eqnarray}
For an observer's time direction $\hat{v}^i$ in $(\Mup, \gup)$ the associated electric and magnetic (Yang-Mills) fields are given by $\hat{E}^\alpha_i = F^\alpha\,_{ij}\hat{v}^j$ and $\hat{B}^\alpha_i = {}^*F^\alpha\,_{ij}\hat{v}^j$. Both covectors have conformal weight $-1$, i.e. $E^\alpha_i = \cfup^{-1}\hat{E}^\alpha_i$.
The fluid density $\hat{\rho}$ and fluid velocity $\hat{u}^i$ have conformal weights $-4$ and $-1$, i.e. $\rho = \cfup^{-4} \hat{\rho} $ and $u^i = \cfup^{-1}\hat{u}^i$. 
In \cite{FriEMYM} and \cite{LueVal radiation fluid} it was shown that near the conformal boundary of asymptotically simple spacetimes the unphysical variables
 $E^\alpha_i, B^\alpha_i, \rho, u_i $ are bounded in $(M,g)$. Thus the physical variables $\hat{E}^\alpha_i, \hat{B}^\alpha_i, \hat{\rho}, \hat{u}^i $ vanish at conformal infinity and $(\Mup, \gup)$ satisfies the cosmic no-hair conjecture.

We now consider a spacetime $(\Mup, \gup)$ with cosmological constant $\cosmolconstup$ satisfying
\begin{eqnarray}
\label{general EFE}
 &&\Gup_{ij} + 6\cosmolconstup \gup_{ij} = \Tup_{ij} = \Tup^{[\scalarfield]}_{ij} + \Tup^{[\mathcal{M}]}_{ij},
\end{eqnarray}  
where $\Tup_{ij} $ has been split into the conformal scalar field part $\Tup^{[\scalarfield]}_{ij} $ and all other trace-free matter components $\Tup^{[\mathcal{M}]}_{ij}$. Due to our assumption of minimal coupling we have $\Dup^i \Tup_{ij}=0$. 

From a mathematical point of view one can also consider an Einstein equation of the form
\begin{equation}
\label{repulsive general EFE}
\Gup_{ij} + 6\cosmolconstup \gup_{ij} =-\Tdown_{ij}.
\end{equation}
We will refer to such matter $\mathcal{M}$ as repulsive matter, e.g. repulsive Einstein-Maxwell or a repulsive radiation fluid. From a physical standpoint such matter is considered unrealistic. However, as we will see below, it will be useful to consider repulsive matter mathematically. Unless otherwise specified we shall always assume that  we are dealing with standard or attractive matter and use \eqref{general EFE}. 
We will refer to the case $\epsilon=-1$ of \eqref{EMtensor conf inv scalar field} as a repulsive conformal scalar field. This terminology is only used to highlight the sign choice in \eqref{EMtensor conf inv scalar field}. Physically speaking it has to be used with caution since, depending on the behaviour of the deriviatives of $\phiup$, $\Tup^{[\scalarfield]} _{ij}$ can be locally negative for $\epsilon=1$ as well as locally positive for $\epsilon=-1$.

We set $\mathbb{T}_{ij}^{[\mathcal{M}]} = \cs^{-2} \Tup_{ij}^{[\mathcal{M}]}$. Using \eqref{EFE RHS}, \eqref{EFE LHS} for $\Tup^{[\scalarfield]}_{ij}$ we can rewrite \eqref{general EFE} in terms of conformal densities to give the following generalisation of \eqref{conformal EFE}:
\begin{equation}
\label{conformal EFE general}
\EMoperator_{ij}[\dualscalarfield, \cosmolconstup] = \epsilon\, \EMoperator_{ij}[\scalarfield, \CouplingConst] + \mathbb{T}_{ij}^{[\mathcal{\mathcal{M}}]}.
\end{equation}

As before we consider switching the roles of $\EMoperator_{ij}[\dualscalarfield, \cosmolconstup]$ and $\EMoperator_{ij}[\scalarfield, \CouplingConst]$. 
Setting $\Tdown_{ij}^{[\mathcal{M}]} = \dualcs^{2} \mathbb{T}_{ij}^{[\mathcal{M}]} $,
the Einstein equation in $(\Mdown, \gdown_{ij})$ takes the form
\begin{equation}
\label{dual EFE matter}
\Gdown_{ij} + 6\CouplingConst \gdown_{ij}=\Tdown_{ij} = \Tdown^{[\dualscalarfield]} _{ij} - \epsilon \Tdown^{[\mathcal{M}]}_{ij} .
\end{equation}
We can see that the additional matter models in  $(\Mup, \gup_{ij})$ and  $(\Mdown, \gdown_{ij})$ are identical, e.g. a Yang-Mills field will lead again to a Yang-Mills field.

For an attractive conformal scalar field ($\epsilon=1$) the additional matter $\mathcal{M}$ changes its characteristic from attractive to repulsive (and vice versa) in the dual solution. This problem is due to the minus sign in \eqref{dual EFE matter} and was already observed in \cite{Bekenstein1974}, where the resulting dual solutions were considered as unphysical. However if we start from a solution whose additional matter $\hat{\cal{M}} $ is repulsive then the dual solution will contain standard matter $\check{\cal{M}} $. We will exploit this aspect later to obtain new solutions with attractive matter.

Note that for the repulsive conformal scalar field ($\epsilon=-1 $) the problem disappears. Both $\Tdown^{[\mathcal{M}]}_{ij} $ and $\Tup^{[\mathcal{M}]}_{ij} $ are either attractive or repulsive. Thus for $\epsilon = -1$ we can have two dual solutions $(\Mup, \gup) $ and $(\Mdown, \gdown)  $, which both describe a repulsive conformal scalar field coupled to attractive trace-free matter.

In summary we have
\begin{theorem}
\label{Duality theorem}
Let $(\Mup, \gup) $ denote a solution of the Einstein field equations with cosmological constant $\Cosmolconstup=6\cosmolconstup $ and a conformal scalar field $\phiup $ with $\CouplingConst$ as coefficient of the self-interaction term in \eqref{EMtensor conf inv scalar field} coupled to additional trace-free matter $\hat{\cal{M}} $. Then the following hold:
\begin{enumerate}
\item There exists a dual solution $(\Mdown, \gdown)  $ with $\gdown_{ij}  =\phiup^2 \gup_{ij} $ with cosmological constant $A=6\CouplingConst$ and a conformal scalar field $\phidown $ with $\cosmolconstup$ as coefficient of the self-interaction term in \eqref{EMtensor conf inv scalar field}. 

\item The conformal scalar field models have the same parameter $\epsilon$.

\item The additional matter $\check{\cal{M}}  $ is of the same type as $\hat{\cal{M}} $.

\item If $\epsilon=1$ then repulsive matter $\hat{\cal{M}} $ gives attractive matter $\check{\cal{M}}$, and vice versa.

\item If $\epsilon=-1$ then the additional matter  components $\hat{\cal{M}} $ and $\check{\cal{M}}$ are either both attractive or both repulsive.
\end{enumerate} 

\end{theorem}
\begin{corollary}
\label{Duality corollary 1}
The dual of an attractive conformal scalar field $(\epsilon=1)$ coupled to a repulsive Yang-Mills  field and a repulsive radiation fluid is an attractive conformal scalar field coupled to an attractive Yang-Mills  field and an attractive radiation fluid.
\end{corollary}

\begin{corollary}
\label{Duality corollary 2}
The dual of a repulsive conformal scalar field $(\epsilon=-1)$ coupled to a Yang-Mills  field and a radiation fluid is an repulsive conformal scalar field coupled to a Yang-Mills  field and a radiation fluid.
\end{corollary}

%
%
%
\noindent Remark: We require $\hat{T}^{[\mathcal{M}]}_{ij}$  to be trace-free, so that $\check{T}^{[\mathcal{M}]}_{ij}$ will be divergence free as well.

\subsection{Relationships between the two physical spacetimes $(\Mup, \gup)$ and $(\Mdown, \gdown)$ and the unphysical manifold $(M,g)$}

Considering $(\Mup, \gup)$ as our physical solution we can see that as long as $\phiup \ne 0$ there is no problem in finding the dual solution $(\Mdown, \gdown)$. In the case $\phiup=0 $ we have $\phidown=\infty$ and the conformal factor relating $\gup$ and $\gdown$ will vanish, see \eqref{3 metrics} and \eqref{relationships between scalar fields} below. Similarly, we are interested in the asympotic behaviour of both spacetimes, i.e the neighbourhood of the set where $\cs$ vanishes. 

In the following we will assume that we can choose a conformal scale $\cv \in \sect[1]$ such that $g_{ij}=\cv^{-2}\cg_{ij}$ is regular throughout $M$. The conformal scales $\cs$ and $\dualcs$ provide a conformal embedding of the physical solutions $(\Mup, \gup)$ and $(\Mdown, \gdown)$ into $(M,g)$. Thus we will consider $(M,g)$ as our unphysical spacetime for both solutions. The corresponding Levi-Civita connections will be denoted $\Dup, \Ddown, \D$ respectively. Analogous notation will be used for quantities related to quantities associated with a particular choice of conformal scale or metric.

The three metrics $g, \,\gup$ and $\gdown$ are defined in terms of the conformal metric $\cg$ by
\begin{equation}
\label{3 metrics defined}
g_{ij} =\cv^{-2} \cg_{ij} , \quad \gup_{ij}  =\cs^{-2} \cg_{ij}, \quad \gdown_{ij} . =\dualcs^{-2} \cg_{ij}
\end{equation}
We define the conformal factors
\begin{equation}
\label{conformal factors}
\cfup=\cs \cv^{-1}  ,\quad\quad \cfdown=\dualcs \cv^{-1} ,\quad\quad \Omegaup=\cs \dualcs^{-1}=\frac{\cfup}{\cfdown}, \quad\quad \Omegadown=\dualcs \cs^{-1}=\frac{\cfdown}{\cfup}.
\end{equation}
Hence the following conformal relationships\footnote{Note that our convention differs from \cite{Tod2013CCC}.} hold between the metrics $g, \,\gup$ and $\gdown$
\begin{equation}
\label{3 metrics}
g_{ij} =\cfup^2 \gup_{ij} , \quad g_{ij} = \cfdown^2 \gdown_{ij} \quad \gdown_{ij}  =\Omegaup^2 \gup_{ij}, \quad \gup_{ij} =\Omegadown^2 \gdown_{ij}.
\end{equation}

The physical scalar fields $\phiup, \phidown $ and the unphysical scalar fields $\Phiup, \Phidown $ are the realisations of the conformal densities $\scalarfield $ and $\dualscalarfield $ in the conformal scales $\cs, \dualcs$ and $\cv$. They are given by
\begin{equation}
\label{definitions of scalar fields}
\phiup=\cs \scalarfield, \quad\quad \phidown=\dualcs \dualscalarfield, \quad\quad 
\Phiup=\cv \scalarfield, \quad\quad \Phidown=\cv \dualscalarfield.
\end{equation}
Recall that the conformal scales $\cs, \dualcs$ and the conformal scalar fields $\scalarfield , \dualscalarfield $ are related by
\begin{equation}
\dualscalarfield = \cs^{-1}, \quad\quad \dualcs=\scalarfield^{-1}.
\end{equation}
Using \eqref{conformal factors} and \eqref{definitions of scalar fields} we find the following relationships between $\phiup, \phidown ,\Phiup$ and $ \Phidown $
\begin{equation}
\label{relationships between scalar fields}
\Phiup = \frac{1}{\cfdown},\quad\quad 
\Phidown = \frac{1}{\cfup}, \quad\quad
\phiup =\cfup \Phiup = \Omegaup, \quad\quad 
\phidown = \cfdown \Phidown = \Omegadown, \quad\quad
\phidown=\phiup^{-1}.
\end{equation}
The energy-momentum tensors in \eqref{general EFE} and \eqref{dual EFE matter}  are related by
\begin{equation}
\label{EM tensor relationship}
\Tdown_{ij} =  \Omegaup^{-2} \Tup_{ij} =  \phiup^{-2} \Tup_{ij}, \quad\quad
\Tup_{ij} =  \Omegadown^{-2} \Tdown_{ij} =  \phidown^{-2} \Tdown_{ij}.
\end{equation}
We recall that a spacetimes $(\Mup,\gup) $ is said to be \emph{weakly asymptotically flat} if there exists a conformally related spacetime $(M,g)$ with $g=\cfup^2 \gup $  satisfying i) $\cfup\vert_{\Mup} >0$ and ii) at $\Scri = \{\cfup=0 \}$ one has $\D_i \cfup \ne 0$. Note minor exceptions to ii) exists when $\cosmolconstup=0$ as there exist isolated points where $\cfup=0 $ and $\D_i \cfup = 0$, but $\D_i\D_j\cfup \ne 0$.

A spacetime $(\Mdown,\gdown) $ has an \emph{isotropic singularity}
\footnote{This is also referred to as a conformally compactifiable singularity \cite{AngTodperfectfluid} or as a conformal gauge singularity \cite{LueTod conf gauge sing}.} if there exists a conformally related spacetime $(M,g)$ with $\gdown = \Xi^2 g$, $\Mdown \subset M $ and a spacelike hypersurface $\Sigma$ where i) $\Xi=0 $, ii) $g$ is a regular metric, iii) $\gdown$ is singular. The hypersurface $\Sigma$ is interpreted as the singularity of $(\Mdown,\gdown) $. Here we mean regular in the sense that the curvature is sufficiently smooth and thus finite at $\Sigma$, while singular means that part of the curvature blows up as we approach $\Sigma$ (strictly speaking $\Sigma$ is not part of $(\Mdown,\gdown) $). Since $g$ is regular at $\Sigma$ the Weyl tensor $C\tensor{ij}{k}{l} $ must be regular at $\Sigma$. Thus the singular behaviour of $\gdown$ must arise from a blow up of the Ricci tensor $\check{R}_{ij} $, respectively the Schouten tensor $\check{\Schouten}_{ij}$ or the energy momentum tensor $\check{T}_{ij} $. 

For a conformal scalar field spacetime the divergence of the scalar field on its own is insufficient for an isotropic singularity. For example at the horizon the extremal Reissner-Nordstr\"om black hole $\Omega=\phi=\infty $, while the spacetime curvature is regular there. Hence the horizon cannot be interpreted as a curvature singularity, or even an isotropic singularity. This example illustrates that for an isotropic singularity we require $\Tdown_{ij} $ to diverge.

\medskip

We proceed by analysing the effects of $\phiup$ vanishing or diverging on a hypersurface $\Sigma$.
\begin{lemma}
\label{Lemma isotropic singularity}
Suppose $\Tup_{ij}$ is given by \eqref{EMtensor conf inv scalar field}. Suppose there exists a hypersurface $\Sigma$ in $(\Mup, \gup)$ where $\phiup$ vanishes smoothly but $\Tup_{ij}\ne 0$. 
Then in the dual spacetime $(\Mdown, \gdown)$ $\phidown$ and $\Tdown_{ij}$ diverge on $\Sigma$ and if $\Sigma$ is a spacelike hypersurface then it represents an isotropic singularity.
\end{lemma}
\proof Set $g=\gup$ and $\Xi = \Omegaup=\phiup $. Then $g$ is regular at $\Sigma$ and $\Xi=0$ there. If $\Sigma$ is spacelike the conditions for an isotropic singularity are satisfied. Equations \eqref{relationships between scalar fields} and \eqref{EM tensor relationship} imply that $\phidown$ and $\check{T}_{ij}$ diverge at $\Sigma$. Since the Weyl tensor is finite  $\Sigma$ represents an isotropic singularity.

\begin{lemma}
\label{Lemma conformal infinity}
Suppose $\Tdown_{ij}$ is given by \eqref{EMtensor conf inv scalar field}. Suppose there exists a hypersurface $\Sigma$ in $(\Mdown, \gdown)$ where $\phidown$ diverges, $(\partial_i \phidown) \phidown^{-2}\ne 0$ and $\Tdown_{ij}$ is finite.
Then the dual spacetime $(\Mup, \gup)$ is weakly asymptotically flat, the hypersurface $\Sigma$ represents part of conformal infinity and $\phiup$ and $\Tup_{ij} $ vanishes on $\Sigma$.
\end{lemma}

\proof The vanishing of $\phiup$ on $\Sigma$ follows directly from \eqref{relationships between scalar fields}. 
Since the curvature in $(\Mdown, \gdown)$ is regular at $\Sigma$ find that $\check{T}_{ij}$ is bounded there and hence $\hat{T}_{ij} = \Omegadown^{-2} \check{T}_{ij}= \phiup^{2} \check{T}_{ij}$ vanishes at $\Sigma$ as well.

Since 
i) $\gdown_{ij}  =\Omegaup^2 \gup_{ij} = \phiup^2 \gup_{ij} $, 
ii) $\Omegaup=\phiup =0$ on $\Sigma$ with $\Omegaup \ne 0$ on $\Mdown \setminus \Sigma$
iii) $\Dup_i \Omegaup = \partial_i(\phidown^{-1})=(\partial_i \phidown) \phidown^{-2} \ne 0$ on $\Sigma$ and 
iv) $\Sigma$ is a regular hypersurface in $(\Mdown, \gdown)$ where the metric $\gdown_{ij}$ is regular, it follows that $\Sigma$ satisfies the definition of conformal infinity for $(\Mup, \gup)$.

Remark: The nature of the conformal boundary of $(\Mup, \gup)$ depends on the cosmological constant $\cosmolconstup$, i.e. the coefficient of the quartic self-interaction term for $\phidown$. 

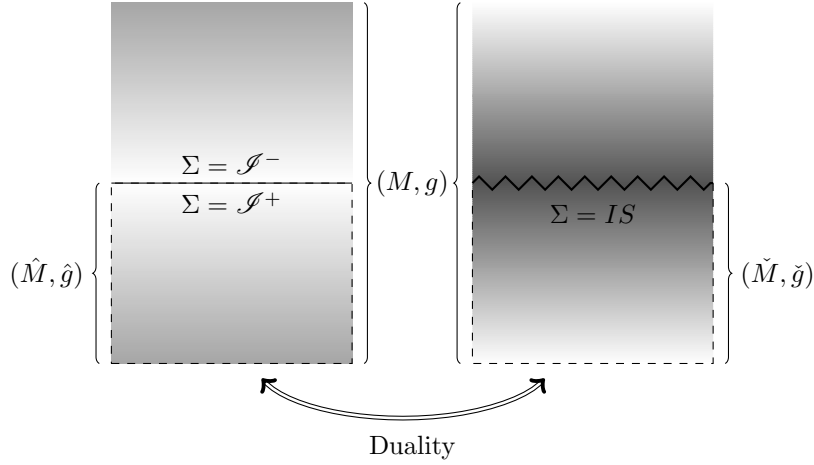
\begin{figure}
\begin{center}
\begin{tikzpicture}[scale=0.8]
\shade[top color=gray!70!white,bottom color=white] (-5,0) -- (-1,0) -- (-1,3) -- (-5,3) -- cycle;
\shade[bottom color=gray!70!white,top color=white] (-5,0) -- (-1,0) -- (-1,-3) -- (-5,-3) -- cycle;
\draw[black] (-5,0) -- (-1,0) ;
\draw [decorate, decoration=brace] {(-5.2,-3) -- (-5.2,0)};
\draw [decorate, decoration=brace] {(-0.8,3) -- (-0.8,-3)};
\draw[dashed] (-1,0) -- (-5,0) -- (-5,-3) -- (-1,-3) -- cycle;
\draw (-5.3, -1.5) node[anchor=east] {$(\hat{M},\hat{g})$};
\draw (0 ,0) node{$(M,g)$};
\draw (-3,0) node[anchor=north] {$\Sigma = \Scri^+$};
\draw (-3,0) node[anchor=south] {$\Sigma = \Scri^-$};

\shade[top color=black!70,bottom color=white] decorate [decoration=zigzag] {(1,0) -- (5,0)} -- (5,-3) -- (1,-3) -- cycle;
\shade[bottom color=black!70,top color=white] decorate [decoration=zigzag] {(1,0) -- (5,0)} -- (5,3) -- (1,3) -- cycle;
\draw[black, thick] decorate [decoration=zigzag] {(1,0) -- (5,0)};
\draw[dashed] (5,0) -- (5,-3) -- (1,-3) -- (1,0);
\draw [decorate, decoration=brace] {(0.8,-3) -- (0.8,3)};
\draw [decorate, decoration=brace] {(5.2,0) -- (5.2,-3)};
\draw (5.3 ,-1.5) node[anchor=west] {$(\check{M},\check{g})$};
\draw (3,-0.2) node[anchor=north] {$\Sigma =IS$};
\draw (0,-4.75) node[anchor=south]{Duality};
\draw [double distance=1pt,<->] (-2.5,-3.25) arc (200:340:2.5cm and 1cm);
\end{tikzpicture}
\end{center}
\caption{\textbf{The conformal duality:} The bridging spacetime $(M,g)$ is depicted (twice). The physical spacetimes
$(\hat{M},\hat{g}) $, a weakly asymptotically flat spacetime, and 
$(\check{M},\check{g})$, a spacetime with an isotropic singularity (IS), can be conformally embedded into $(M,g)$. The surface $\Sigma$ represents null infinity, respectively the isotropic singularity (zigzagged line). The strength of each (physical) energy-momentum tensor is indicated by the strength of the shading.
Note the upper halves also describes two dual physical spacetimes.}
\label{Duality graphic}
\end{figure}

\medskip
For the CCC-scenario we would like $\Sigma$ to describe both conformal infinity and the isotropic singularity. However, neither conformal infinity nor the isotropic singularity are part of their physical spacetime, while the above lemmas assume explicitly that $\Sigma$ is a regular hypersurface in the respective physical spacetime. Below we will show that working with a bridging spacetime $(M,g)$ and making suitable assumptions, $\Sigma$ can describe conformal infinity for $(\Mup, \gup)$ and the isotropic singularity for $(\Mdown, \gdown)$ simultaneously.
\begin{lemma}
\label{Lemma CCC scenario}
Suppose $(\Mup, \gup)$ is a weakly asymptotically flat spacetime with a conformal scalar field $\phiup$ and energy momentum tensor \eqref{EMtensor conf inv scalar field}. 
Suppose $(\Mup, \gup)$ is conformally embedded into $(M,g)$ and the conformal boundary of $(\Mup, \gup)$ is described by a spacelike hypersurface $\Sigma$ in $(M,g)$.
Suppose that $\Phiup=\phiup \cfup^{-1} $ vanishes on $\Sigma$, while $T_{ij} = \cfup^{-2} \Tup_{ij}$ is finite, but non-vanishing there. Then in the dual spacetime $(\Mdown, \gdown)$ the hypersurface $\Sigma$ represents an isotropic singularity. (See Figure \ref{Duality graphic})
\end{lemma}
\proof Combining \eqref{3 metrics} and \eqref{relationships between scalar fields} we find $\gdown_{ij} = \cfdown^{-2}g_{ij} = \Phiup^2 g_{ij} $. Since $T_{ij} $ is non-zero we find that $ \Tup_{ij} = \cfup^{2}T_{ij} $ vanishes while $ \Tdown_{ij} = \cfdown^{2}T_{ij} $ diverges. Furthermore $(M,g)$ has  regular Weyl curvature and hence $(M,g)$ and $(\Mdown, \gdown)$ satisfy the requirements for $\Sigma$ to represent an isotropic singularity for $(\Mdown, \gdown)$.

\medskip

\medskip
Remark: 
In Lemma \ref{Lemma isotropic singularity} and \ref{Lemma CCC scenario} the requirement that $\Sigma$ be spacelike has been chosen here in order to fit the definition of isotropic singularities in \cite{GW-CQG1985}, which uses the level set $\Sigma=\{T=0\}$ of a time function $T$. The condition can be dropped if one allows for isotropic singularities along null or timelike hypersurfaces.
\begin{lemma}
\label{Lemma CCC scenario with matter}
The above three lemmas hold if the conformal scalar field is coupled to trace-free matter, in particular Yang-Mills fields and radiation fluids. 
\end{lemma}
\proof Recall that
$\Tup_{ij} = \Tup^{[\varphi]}_{ij}+\Tup^{[\mathcal{M}}_{ij}$ implies $T_{ij} = T^{[\varphi]}_{ij} + T^{[\mathcal{M}}_{ij}$ where each part rescales individually by $\Xi^{-2}$. In particular, recall that the fields $E^\alpha_i, B^\alpha_i, \rho, u^i $ have conformal weights $-1,-1,-4,-1$ and their values in the spacetime $(M,g)$ are finite.

Lemma 1: Again set $\cv=\cs$, so that $\cfup=1$ and $(M,g)=(\Mup, \gup)$. It follows that $\Omegaup=\phiup=\cfdown^{-1}=\phidown^{-1}$. Thus, $\Omegaup=0$ and $\phidown=\infty$ at $\Sigma$. Since $\Tup_{ij}\ne 0 $, \eqref{EM tensor relationship} implies that $\Tdown_{ij}$ diverges while the Weyl tensor is finite. Hence $\Sigma$ represents an isotropic singularity. In particular, the fields $E^\alpha_i, B^\alpha_i, \rho, u^i $, which have same finite values in $(M,g)$ and $(\Mup, \gup)$, rescale to $\check{E}^\alpha_i = \Omegaup^{-1}E^\alpha_i, \check{B}^\alpha_i = \Omegaup^{-1}B^\alpha_i, \check{\rho} = \Omegaup^{-4}\rho$. Thus any field that does not vanish in $(M,g)=(\Mup, \gup)$ will diverge in $(\Mdown, \gdown)$. 

Lemma 2: The proof goes directly through as before with $\Tup_{ij}, \hat{E}_i , \hat{B}_i , \hat{\rho}$ each vanishing at $\Sigma$.

Lemma 3: Once more $ \Tup_{ij} = \cfup^{2}T_{ij} $ vanishes while $ \Tdown_{ij} = \cfdown^{2}T_{ij} $ diverges at $\Sigma$, which hence represents an isotropic singularity in $(\Mdown, \gdown)$. 

\medskip
Remark: Lemma \ref{Lemma CCC scenario with matter} works for $\epsilon=\pm 1$ as well as standard or repulsive matter being chosen.

\section{Spacetimes with a conformal scalar field}
\label{WAFST}

In this section we will prove the existence of spacetimes with a conformal scalar field that are either weakly asymptotically flat or contain an isotropic singularity. Our approach uses the conformal Einstein field equations (CEFE). As Friedrich showed, a key feature of the CEFE for vacuum \cite{FrideSitter}, \cite{FriJGP} or trace-free matter \cite{FriEMYM} is that a solution of the CEFE on $(M,g)$ implies a solution of the related Einstein field equations on each connected component of the subset $\{ \Theta \ne 0\} $ of $(M,g)$. The method has been extended to spacetimes containing Einstein-Maxwell-Yang-Mills \cite{FriEMYM}, radiation fluid \cite{LueVal radiation fluid} or conformal scalar field \cite{Huebner}. In particular, one can formulate a Cauchy problem for each of these CEFE. In order to prove existence and uniqueness for the CEFE one shows that the CEFE can be transformed into a first order symmetric hyperbolic system and that the constraints are propogated. As we will discuss in the next section the method can be directly extended to spacetimes containing a combination of the above mentioned matter fields in a minimally coupled form. In particular, one can prove the following theorem
\begin{theorem}
\label{Existence Theorem}
Let $\Sigma$ be an initial surface with initial data for the CEFE of a standard conformal scalar field $(\epsilon=1)$ minimally coupled to Einstein-Maxwell-Yang-Mills and a radiation fluid. Suppose that on $\Sigma$ the initial conformal scalar field satisfies $-1<\Theta_*\phi_* < 1$. Then in a neighbourhood $U$ of $\Sigma$ there exists a solution to the CEFE. In each connected component of $U \setminus \{\theta=0\}$ the solution gives rise to a physical spacetime whose matter model is a conformal scalar field minimally coupled to Einstein-Maxwell-Yang-Mills and a radiation fluid.
\end{theorem}

We are free to set $\Theta_*=1 $ and $\phi_*=0$ across $\Sigma$. Thus we find
\begin{corollary} 
\label{Existence Corollary}
There exist spacetimes with a conformal scalar field minimally coupled to Einstein-Maxwell-Yang-Mills and a radiation fluid for which the conformal scalar field vanishes on a regular spacelike hypersurface $\Sigma$.
\end{corollary}
Remark: There is no restriction in Theorem \ref{Existence Theorem} and Corollary \ref{Existence Corollary} on the Einstein-Maxwell-Yang-Mills fields and the radiation fluid. They can be attractive or repulsive.

\subsection{The CEFE for the coupled system}
As outlined above we will use the unphysical momentum tensor $T_{ij} = T^{[\varphi]}_{ij}+T^{[EMYM]}_{ij}+T^{[rad]}_{ij}$ with $\D^i T^{[\varphi]}_{ij}=0$, $\D^i T^{[EMYM]}_{ij}=0$ and $\D^i T^{[rad]}_{ij}=0$ implying $\D^i T_{ij}=0$.

The CEFE for Einstein-Maxwell-Yang-Mills, conformal scalar field and radiation fluid were discussed in detail in \cite{FriEMYM}, \cite{Huebner}, \cite{LueVal radiation fluid} respectively. The general details can be found in these references. Here we will only focus on the essential points.

The variables used in the CEFE can be split into geometric variables (frame fields, connection coefficients, Schouten tensor and Weyl curvature) and matter variables (conformal scalar, Yang-Mills fields, fluid density and fluid velocity). As shown in \cite{FriEMYM}, \cite{Huebner}, \cite{LueVal radiation fluid} one needs to include derivatives of the matter variables in order to obtain an overall first order system. In the following we will refer to these derivatives as matter fields as well. The prinicipal part of the CEFE for the geometric variables is identical to the vacuum case and due to the minimal coupling of the matter models the principal part for the matter variables is a disjoint combination of the equations for the individual models.

In the case of a conformal scalar field two equations require particular attention. In terms of the metric $g=\cfup^2 \gup$
\footnote{In \cite{Huebner} the physical variables are denoted with a tilde, while unphysical variables have no marker. The equations from \cite{Huebner} presented her have been adapted to our notation, so that $\tilde{\phi}$, $\phi$ and $\Omega$ have been replaced by $\phiup$, $\Phiup$ and $\cfup $}
 and its Levi-Civita connection $\D$ they take the form
\begin{eqnarray}
\label{div Weyl}
\D_k d\tensor{ij}{k}{l} &=& t_{ijl} \\
\label{Cotton-York}
2 \D_{[i} \Schouten_{j]l} &=& d\tensor{ij}{k}{l} d_k + \cfup t_{ijl}
\end{eqnarray}
where $d\tensor{ij}{k}{l} = \cfup^{-1} C\tensor{ij}{k}{l} $, $d_i=\D_i \cfup $ and 
 $t_{ijl} =  2 \cfup^{-1} \Dup_{[i} \Schoutenup_{j]l}$. For trace-free matter models $t_{ijl} $ can be expressed in terms of $T_{ij}$ and $\D$ as follows
\begin{equation}
\label{t from T equation}
t_{ijl} = \cfup \D_{[i} T_{j]l} + 3 d_{[i} T_{j]l} - g_{l[i} T_{j]k} d^k.
\end{equation}
Since \eqref{t from T equation} is linear in $T_{ij}$ we have $t_{ijk} = t^{[\varphi]}_{ijk}+t^{[EMYM]}_{ijk}+t^{[rad]}_{ijk} $. The problem that arises in the case of a spacetime with a single conformal scalar field is that $\Theta \D_{[i} T_{j]l} $ and thus $t_{ijl}$ gives rise to $\D_{[i} \Schouten_{j]l} $ appearing on the RHS of \eqref{div Weyl} and \eqref{Cotton-York}. In this form the CEFE cannot be rendered into a first order system. As shown in \cite{Huebner}, one expands $t_{ijl}$ to give 
$$
t_{ijl} = 2 \cfup \Phiup^2 \D_{[i} \Schouten_{j]l} + \mathrm{l.o.t.}\,\,.
$$
Here lower order terms (l.o.t.) mean those terms which are polynomial combinations of the variables but contains no explicit derivatives. Now substituting \eqref{Cotton-York} gives
\begin{equation}
\label{m tensor}
t_{ijl} = m_{ijl} := (1-\cfup^2 \Phiup^2)^{-1} \left( \cfup \Phiup^2 d\tensor{ij}{k}{l} d_k + \mathrm{l.o.t.} \right).
\end{equation}
Thus $m_{ijl}$ contains no explicit derivatives and \eqref{div Weyl} and \eqref{Cotton-York} can be rewritten as 
\begin{eqnarray}
\label{div Weyl recast}
\D_k d\tensor{ij}{k}{l} &=& m_{ijl} \\
\label{Cotton-York recast}
2 \D_{[i} \Schouten_{j]l} &=& d\tensor{ij}{k}{l} d_k + \cfup m_{ijl}
\end{eqnarray}
leading to the desired form of the CEFE. Observe that $t_{ijl}$, and hence $m_{ijl}$, contains $\D_{[i} T^{[EMYM]}_{j]l} $ and $\D_{[i} T^{[rad]}_{j]l} $. Since the derivatives of the matter fields have been introduced as additional variables, one can replace any derivatives in $\D_{[i} T^{[EMYM]}_{j]l} $ and $\D_{[i} T^{[rad]}_{j]l} $ in terms of these additional variables. Therefore in the minimally coupled problem analysed here, the tensor $m_{ijl}$ thus contains matter variables but no explicit derivatives. The remainder of the argument follows the proof in \cite{Huebner}. 
As in the case of a single conformal scalar field one has to restrict the system to the region where $1-\cfup^2\Phiup^2>0 $
\footnote{In \cite{Huebner} the conditon is $1-\frac{1}{4}\Omega^2\phi^2>0 $ since $\phi$ differs by a factor from $\Phiup$.}
, i.e. $\vert \phiup \vert =\vert \cfup \Phiup \vert < 1$, in order for the CEFE and their subsequent hyperbolic reduction to form a regular system of PDEs. In the following we restrict ourselves to  the case $-1<\cfup \Phiup < 1$. This still allows us to work in the asymptotic region. The arguments for the existence and uniqueness presented in \cite{Huebner} essentially follow \cite{FriEMYM} and carry through to the setting given here. This proves Theorem \ref{Existence Theorem}. Observe that both attractive and repulsive versions of Einstein-Maxwell-Yang-Mills and radiation fluid are permitted, since the repulsive cases only introduce a factor of $-1$ for some of the variables but do not affect the principal part of PDE system and hence Theorem \ref{Existence Theorem} holds as well.

Remark: If $\cfup \ne 0$  everywhere on $\Sigma$, then $\Sigma$ is a regular hypersurface  and we can use the conformal freedom to work with $\cfup=1$ everywhere. The CEFE simplify to the problem in the physical spacetime and the restriction becomes $1-\phiup^2 >0$. The local solution then describes a regular region of the physical spacetime away from the conformal boundary.

\subsection{The CEFE for a conformal scalar field with $\epsilon=-1$}
The derivation of the CEFE is almost identical for the case  $\epsilon=-1$. The key difference rests on the fact that in \eqref{m tensor} the factor is now $(1+\cfup^2\Phiup^2 )^{-1}$, which is regular for all values of $\phiup=\cfup \Phiup$ as long as neither $\cfup$ or $\Phiup$ diverge. Following through the remainder of the argument one then finds that the CEFE and their hyperbolic reduction are regular and that for $\epsilon=-1$ no restriction on $\phiup=\cfup \Phiup$ is required. We thus have the following version of Theorem \ref{Existence Theorem} for a conformal scalar field with $\epsilon=-1$
\begin{lemma}
\label{Existence Theorem repulsive version}
Let $\Sigma$ be an initial surface with initial data for the CEFE of a conformal scalar field with $\epsilon=-1$, which is minimally coupled to Einstein-Maxwell-Yang-Mills and a radiation fluid. Then a solution to the CEFE exists in a neighbourhood of $\Sigma$, which gives rise to a physical spacetime whose matter model is a conformal scalar field with $\epsilon=-1$ minimally coupled to Einstein-Maxwell-Yang-Mills and a radiation fluid.
\end{lemma}

\subsection{Asymptotically flat spacetimes with a conformal scalar field}

For the de Sitter-like case, $\lambda <0$, one can setup a Cauchy problem for the CEFE on the spacelike hypersurface $\Sigma$ representing $\Scri$. This is referred to as the initial value problem at null infinity. The solution, which exists in a neighbourhood of $\Scri$, generates two physical solutions. One solution lies to the past and has a future null infinity, the other one lies to the future with a past null infinity. These two solutions are then future (past) geodesically complete and weakly asymptotically flat.
  
The construction of such spacetimes is discussed below.
Let $\Sigma$ denote the initial surface and $\{n, e_a\}$ form a $g$-orthonormal frame, where $n^i$ is the normal of $\Sigma$. In the following the letters $a,b,c,d$ are reserved for spatial indices and take values $1,2,3$.
Let $\cosmolconstup<0$ denote the cosmological constant and $\cosmolconstdown$ the coefficient of quartic self-interaction term. Recall that $d_i=\D_i \cfup$, so that $d_0= \D_n \cfup$. Let $h_{ab}$ denote the 3-metric induced on $\Sigma$. Furthermore, $d\tensor{ij}{k}{l}=\cfup^{-1} C\tensor{ij}{k}{l}$ and we let  $d_{ab} = d_{a0b0} $ and $d_{abc}=d_{a0bc} $. Then the rescaled electric and magnetic Weyl tensor are given by
$\mathcal{E}_{ab} = d_{ab} $, $\mathcal{H}_{ab} = d_{a0cd}\epsilon\tensor{b}{cd}{} $ \footnote{$\epsilon_{bcd} = \epsilon_{0bcd}$ is the 3 dimensional volume form of $h_{ab}$.}

\subsubsection{The conformal constraints}

In order to find the initial values for the CEFE  at $\Scri$ we analyse the conformal constraint equations at $\Scri$, i.e. $\cfup=0$ and $d_a=D_a \cfup = 0$. 
It was shown in \cite{FriJGP}, \cite{FriEMYM} that the conformal constraints at $\Scri$ can be solved. 
The initial data for the geometric variables is given by
\begin{subequations}
\label{Initial data}
\begin{eqnarray}
&&  d_0  = \sqrt{-\lambda/3}, \quad 
s= \sqrt{-\lambda/3}\,t, \quad
\chi_{ab} = -t h_{ab},  \\
&& L_a = D_a t, \quad
L_{ab}= l_{ab}, \quad \mathcal{H}_{ab} = -\sqrt{-3/\lambda} B_{ab}\\
\label{electric Weyl constraint}
&& D^a \mathcal{E}_{ab}= \sqrt{-\lambda/3} \,\,T_{b0}. 
\end{eqnarray}
\end{subequations}
where $t$ is a smooth real function, $h_{ab}$ is a 3-metric on $\Sigma$, $\mathcal{E}_{ab}$ is a symmetric trace-free tensor and $B_{ab}$ and $y_{abc}$ are the Bach and the Cotton-York tensor of the 3-metric $h_{ab}$
\[
B_{cd} \epsilon\tensor{}{d}{ab} = - y_{abc} = D_b l_{ac} - D_a l_{bc}.
\]
Observe that $\Schouten_{00}$ or equivalently $\Schouten=\Schouten\tensor{k}{k}{}$ have not been determined by the initial data. $L$ can be chosen freely as a gauge-source function for the conformal factor $\cfup$. A convenient choice is a constant $L$.

For the conformal scalar field, the Einstein-Maxwell-Yang-Mills case and the radiation fluid the RHS of \eqref{electric Weyl constraint} is given by 
\begin{subequations}
\begin{eqnarray}
\label{T0b conf scalar field}
T^{[\phi]}_{0b} &=& 4 \Xi D_b \Phiup  - 2 \Phiup D_b \Xi - 2 \chi\tensor{b}{c}{} D_c \Phiup + 2 \Phiup^2 \Schouten_{0b},\\
\label{T0b EMYM}
T^{[EMYM]}_{b0} &=& F^\alpha\,_{bc} F_\alpha\,\tensor{}{c}{0}, \\
\label{T0b radiation fluid}
T^{[rad]}_{b0} &=& \frac{4}{3}\rho \sqrt{1-u_c u^c} u_b,
\end{eqnarray}
\end{subequations}
where $\Xi = \D_n \Phiup$. For the repulsive analogues multiply the RHS by $-1$.

If we set $\Phiup_*=0$, and hence $(D_a \Phiup)_*=0$, then \eqref{T0b conf scalar field} reduces to  $(T^{[\phi]}_{0b})_* = 0$. If the Maxwell field at $\Scri$ is either purely electric or purely magnetic then $(T^{[EMYM]}_{0b})_* = 0$. If the radiation fluid is orthogonal to $\Scri$ then $(u_b)_*=0$ and $(T^{[rad]}_{0b})_* = 0$.

\medskip
Thus for the Cauchy problem at $\Scri$ the free geometrical data is given by a scalar function $t$, a 3-metric $h_{ab}$ and a symmetric trace-free tensor $\mathcal{E}_{ab}$ satisfying \eqref{electric Weyl constraint}, which itself depends on \eqref{T0b conf scalar field}-\eqref{T0b radiation fluid}. This geometrical initial data is supplemented with solutions to the conformal constraint equations for the matter variables. Overall, we then obtain initial data for the Cauchy problem at null infinity. Using Theorem \ref{Existence Theorem} and Lemma \ref{Existence Theorem repulsive version} we can thus develop  a large class of weakly asymptotically flat spacetimes with a conformal scalar field coupled to Yang-Mills fields and radiation fluids.

In particular we find that
\begin{corollary} 
There exists a class of weakly asyptotically flat spacetimes $(\Mup,\gup)$ for which the unphysical conformal scalar field vanishes at null infinity, i.e. $\cfup_*=0$ and $\Phiup_*=0$.
\end{corollary}

\subsection{Isotropic singularities in spacetimes with a conformal scalar field}

We start from initial data at null infinity which describes a conformal scalar field coupled to repulsive Einstein-Maxwell-Yang-Mills and a repulsive radiation fluid. Using Theorem \ref{Existence Theorem} we obtain the corresponding physical solution $(\Mup, \gup) $. By Corollary \ref{Duality corollary 1} the dual solution $(\Mdown, \gdown) $ describes a conformal scalar field coupled to attractive Einstein-Maxwell-Yang-Mills and an attractive radiation fluid. Applying Lemma \ref{Lemma CCC scenario with matter} we find that $(\Mdown, \gdown) $ has an isotropic singularity. Since the Weyl tensor vanishes at the conformal infinity of $(\Mup, \gup) $ it must vanish at the isotropic singularity of $(\Mdown, \gdown) $.
We thus have the following result.
\begin{theorem}
\label{Theorem strong WCH}
Let $\Sigma$ be a spacelike hypersurface. Suppose on $\Sigma$ we are given initial data at null infinity for the CEFE of a conformal scalar field coupled to repulsive Einstein-Maxwell-Yang-Mills and a repulsive radiation fluid. Then there exists a solution $(\Mdown, \gdown) $ to the Einstein field equations with conformal scalar field coupled to Einstein-Maxwell-Yang-Mills and a radiation fluid that satisfies the strong Weyl curvature hypothesis. In other words, $(\Mdown, \gdown) $ has an isotropic singularity at which the Weyl curvature vanishes.
\end{theorem}
The rescaled Weyl tensor need not vanish at $\Sigma$ and hence the Weyl tensor will be non-zero away from $\Scri$. Hence the above spacetimes provide a new class of spacetimes which satisfy the strong Weyl curvature hypothesis but are not conformally flat. Therefore Theorem \ref{Theorem strong WCH} extends the results in \cite{AngTodVlasov}. Moreover, it shows that despite the apparent fine tuning required by the strong Weyl curvature hypothesis there exists a large class of spacetimes satisfying the hypothesis. 

The work in \cite{AngTodperfectfluid}, \cite{AngTodVlasov}, \cite{Tod-IS deSitter} showed the existence of spacetimes with isotropic singularities satisfying the weak Weyl curvature hypothesis. Suppose we start with initial data for the CEFE of a conformal scalar field coupled to repulsive Einstein-Maxwell-Yang-Mills and a repulsive radiation fluid on a regular hypersurface of $(\Mup, \gup) $, i.e. $\cfup \ne 0$. In a generic spacetime the Weyl tensor will be non-zero away from $\Scri$, so that it is finite and non-vanishing on $\Sigma$. Combining Corollary \ref{Existence Corollary} with Lemma \ref{Lemma isotropic singularity} we get
\begin{corollary}
There exist solutions $(\Mdown, \gdown) $ to the Einstein field equations with conformal scalar field coupled Yang-Mills fields and radiation fluids that have an isotropic singularity at which the Weyl curvature does not vanish identically and that hence satisfy the weak Weyl curvature hypothesis.
\end{corollary}

\subsection{Stability}

In \cite{FriEMYM}, \cite{LueValEMYM} and \cite{LueVal radiation fluid} the stability of Einstein-Maxwell-Yang-Mills and of radiation fluid spacetimes were studied. In particular the stability of de Sitter space in the class of Einstein-Maxwell-Yang-Mills spacetimes and the stability of FLRW in the class of radiation fluids were proven. For the conformal scalar field this aspect of stability was briefly addressed in Theorem 4 of \cite{Huebner}. Since the CEFE can be written as a regular first order symmetric hyperbolic system on can apply theorems by Kato 
\cite{Kato75} to prove stability. Thus given a regular (reference or background) solution of the CEFE, one can prove that it is stable under small perturbations. 

We observe that for conformal scalar fields and Einstein-Maxwell-Yang-Mills we can consider vacuum or electro-vacuum spacetimes as a background spacetime. In particular, we can use the  radiative electro-vacuum spacetimes ($\cosmolconstup=0$) as reference solutions for the CEFE and consider small perturbations in the family of spacetimes containing conformal scalar fields minimally coupled to Einstein-Maxwell-Yang-Mills fields. Note that we cannot include perturbations by radiation fluids for these spacetimes since the formulation of the CEFE in \cite{LueVal radiation fluid} require that the unphysical fluid density never vanishes. However we can set FLRW with $\cosmolconstup<0$ and $k=1$ as our reference time (see \cite{LueVal radiation fluid}) and study small perturbations with respect to all three matter models.
We can thus prove that
\begin{theorem}
$\quad$

\noindent a) The region near future null infinity of radiative electromagnetic spacetimes is stable against small perturbations within the class of spacetimes containing conformal scalar fields minimally coupled to Einstein-Maxwell-Yang-Mills fields.

\noindent b) The region near future null infinity of de Sitter space is stable against small perturbations within the class of spacetimes containing conformal scalar fields minimally coupled to Einstein-Maxwell-Yang-Mills fields.

\noindent c) The region near future null infinity of the radiation fluid FLRW spacetime with $\cosmolconstup<0$ and $k=1$ is stable against small perturbations within the class of spacetimes containing conformal scalar fields minimally coupled to Einstein-Maxwell-Yang-Mills fields and radiation fluids.
\end{theorem}
For the classes of spacetimes in part b) and c) one can consider perturbations of the initial values at null infinity. Since the conformal scalar field $\Phiup $ in the reference solutions vanishes at null infinity, we can consider the subclass of perturbations for which the initial data at conformal infinity satisfies $\Phiup_* =0$ and $\Xi_* = (\nabla_n \Phiup) \vert_\Sigma \ne 0 $. Every spacetime generated by this subclass of initial data is weakly asymptotically flat and has a dual solution which contains an isotropic singularity. Note that in the reference spacetimes chosen in b) and c) are not part of the above subclass since both reference sapcetimes satisfy $\Phiup=0$ everywhere.

Remark: All the reference spacetimes used above have a vanishing conformal scalar field. While explicit solutions with conformal scalar field exist, they typically consider either $\cosmolconstup=0$ or $\CouplingConst=0$. However, it should be possible to generalise some of these solutions under suitable symmetry assumptions.

\section{A proposal for the CCC-scenario}
\label{CCC-proposal}

The CCC-scenario proposes that future conformal infinity of $(\Mup, \gup)$ and the past isotropic singularity of $(\Mdown, \gdown)$ are identified at the spacelike hypersurface $\Sigma$. This requires $\cosmolconstup<0 $ in order to guarantee a spacelike future null infinity $\Scri^+$ (see remarks above). The results in the previous sections allow us to construct explicit solutions to this scenario. 

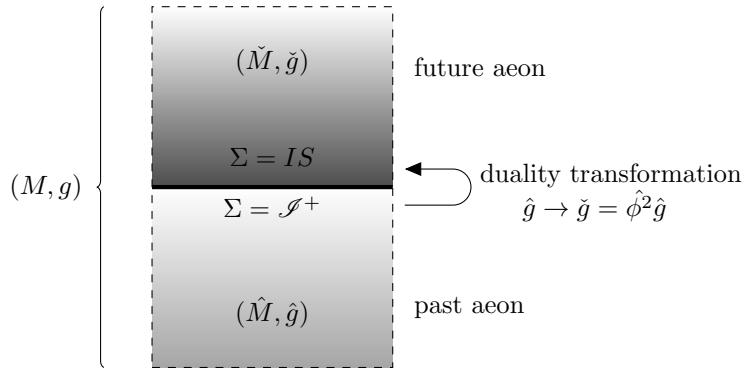
\begin{figure}
\begin{center}
\begin{tikzpicture}[scale=0.8]
\shade[bottom color=black!70,top color=white] {(4,0) -- (0,0)} -- (0,3) -- (4,3) -- cycle;
\shade[bottom color=gray!70!white,top color=white] (0,0) -- (4,0) -- (4,-3) -- (0,-3) -- cycle;
\draw[dashed] (4,-3) -- (4,3) -- (0,3) -- (0,-3) -- cycle;
\begin{scope}[line width = 2pt]
\draw[ultra thick]  {(0,0) -- (4,0)};
\end{scope}
\draw (2,2.5) node[anchor=north] {$(\check{M},\check{g})$};
\draw (2,0.2) node[anchor=south] {$\Sigma =IS$};
\draw (2, -2.5) node[anchor=south] {$(\hat{M},\hat{g})$};
\draw (2,0) node[anchor=north] {$\Sigma = \mathscr{I}^+$};
\draw (4.2, 2) node[anchor=west] {future aeon};
\draw (4.2, -2) node[anchor=west] {past aeon};
\draw [decorate, decoration=brace] {(-0.8,-3) -- (-0.8,3)};
\draw (-1, 0) node[anchor=east] {$(M,g)$};
\draw (4.2,-0.3) -- (5,-0.3);
\draw (5,-0.3) arc (-90:90: 0.3cm and 0.3cm);
\begin{scope}[>=triangle 45]
\draw [->] (5,0.3) -- (4.2,0.3);
\end{scope} 
\draw (5.3,0.2) node[anchor= west] {duality transformation};
\draw (6,-0.3) node[anchor= west] {$\hat{g} \to \check{g}=\hat{\phi^2} \hat{g} $};
\end{tikzpicture}
\end{center}
\caption{\textbf{Proposal for CCC-scenario:} The past aeon is represented by the weakly asymptotically flat spacetime $(\hat{M},\hat{g}) $ for which $\Sigma$ represents future null infinity. The future aeon is represented by the spacetime $(\check{M},\check{g})$ with its past isotropic singularity. The transition between the two aeons is implemented by rescaling by $\hat{\phi}^2$ (duality transformation) at $\Sigma$ (big bang). The shading indicates the strength of the (physical) energy-momentum tensor.}
\label{CCC graphic}
\end{figure}

Recall from our discussion above that the solution of the CEFE with initial data at null infinity gives rise one unphysical solution $(M,g)$. From $(M,g)$ we obtain two separate physical solutions, one to the past and one to the future of $\Sigma$. Denote the solution to the past $(\Mup, \gup)$. By Lemma \ref{Lemma CCC scenario with matter} the solution to the future has itself a dual solution that contains an isotropic singularity with vanishing Weyl curvature. Denote this dual solution $(\Mdown, \gdown)$. Then the two spacetimes $(\Mup, \gup)$ and $(\Mdown, \gdown)$ are joint at the surface $\Sigma$, which in $(\Mup, \gup)$ represents future null infinity and for $(\Mdown, \gdown)$ represents the isotropic singularity (Figure \ref{CCC graphic}). The unphysical spacetime $(M,g)$ plays the role of the bridging spacetime. The spacetimes $(\Mup, \gup)$ and $(\Mdown, \gdown)$ satisfy the criteria of the CCC-scenario.

In order to guarantee that the next aeon ends in a spacelike future null infinity, we need to make sure that the coefficient $\CouplingConst$ of the quartic self-interaction term in this aeon, which becomes the cosmological constant of the next aeon, is negative.

\subsection{Discussion of the proposal}

We should first of all highlight that this proposal is close in its nature to observations made in \cite{Tod ERE}, \cite{Tod2013CCC}. In \cite{Tod ERE} a single radiation fluid in $(\Mup, \gup)$ leads to a radiation fluid coupled to a conformal scalar field in $(\Mdown, \gdown)$. However, the conformal factor is known only implicitly through wave equations.
In contrast, our proposal uses a conformal scalar field in both $(\Mup, \gup)$ and $(\Mdown, \gdown)$ and the duality provides a mathematical mechanism for fixing the conformal factor relating the two physical spacetimes, irrespective of the chosen bridging metric. As seen above this has allowed for the generation of a large class of spacetimes satisfying the CCC-scenario.

If we are dealing with a single conformal scalar field as our matter then we get a satisfactory transition between $(\Mup, \gup)$ and $(\Mdown, \gdown)$ since both spacetimes are described by the same type ($\epsilon=\pm 1$) of conformal scalar field. However, if we couple the conformal scalar field with $\epsilon=1$ to Einstein-Maxwell-Yang-Mills and a radiation fluid then one of the two spacetimes contains standard attractive matter fields while the other sapcetime contains repulsive matter fields. This does not seem satisfactory in light of the fact that such repulsive matter is considered unphysical. 

However, one could argue that our own observations of attractive matter fields satisfying the dominant energy condition only give us information about our current aeon. We cannot infer that the same should hold in the previous or the subsequent aeon (assuming they exist). It is unclear whether the aeon with repulsive matter develops any pecularities or whether gravitational clumping could be largely overcome by the repulsive matter and the presence of a de Sitter-like cosmological constant, leading to an expanding and potentially fairly homogeneous spacetime which is weakly asymptotically flat to the future. 

If one considers the case $\epsilon =-1 $ then both $(\Mup, \gup)$ and $(\Mdown, \gdown)$ contain standard attractive Einstein-Maxwell-Yang-Mills fields, radiation fluids and possibly additional attractive conformal scalar fields. Hence the matter contents of subsequent aeons satisfies the same physical properties. The second advantage of $\epsilon =-1 $ is that the CEFE and the resulting Einstein field equations hold everywhere as $(1+\phi^2)$ never vanishes. What is unclear for this scenario is whether there exists any physical process which is described by \eqref{EMtensor conf inv scalar field} with $\epsilon =-1 $ and could thus account for triggering the switch from one aeon to the next.

\subsubsection{The problem with $\phi=\pm 1$ for $\epsilon =1 $}

Since $\phiup \to -\phiup$ preserves \eqref{EMtensor conf inv scalar field}, we only need to consider $\phiup=1$. The factor $(1-\phiup^2)^{-1}$ diverges in that case. We observe $ \phiup=1 \implies \phidown=1$ so that the dual solution faces the same problem in the same location. Some authors refer to this behaviour as a singularity. This viewpoint may be justified when talking about the CEFE degenerating as a PDE system. For this PDE problem one has to consider other methods for determining whether a solution can be extended past $\phiup=1$. However from a spacetime point of view, $\phiup=1$ does not imply a curvature singularity. The factor in square brackets in \eqref{EMtensor scalar field only} may vanish as well so that the quotient can give a finite limit. If the metric is sufficiently regular then its curvature is bounded, in particular the Einstein tensor is. However this implies in turn a finite energy momentum tensor. The Reissner Nordstr\"om black hole is an illustrative example for this. $ \phiup=1$ at $R=m$, which is a regular location in the exterior region where the curvature and energy momentum tensor are bounded. 

\section{Discussion and comments}

A natural question to ask is whether the above stability result for the weakly asymptotically flat end leads to a stability result for the dual spacetimes with isotropic singularities. We argue for caution in this case. Firstly, the reference spacetimes themselves have no dual solution (since $\phiup=0 $ everywhere) against which to make a comparison. Secondly, the stability proof tells us that the unphysical spacetimes $(M,g)$ are close. So are the weakly asymptotically flat solutions $(\Mup, \gup)$ since $\cfup << 1$ near $\Sigma$. However, in $(\Mdown, \gdown)$ small unphysical perturbations are increasingly magnified in the physical quantities closer to the singularity, since $\cfdown >>1$ near $\Sigma$. 

Moreover, suppose we start from an initial surface close null infinity and study small perturbations. Then our stability results guarantee that the two surfaces where $\Phiup$  and $\cfup$ vanish are close to each other, but they need not coincide. Thus the null infinity of  $(\Mup, \gup)$ and the isotropic singularity of $(\Mdown, \gdown)$ may no longer coincide. Note that Lemma \ref{Lemma isotropic singularity} still assures existence of the singularity in $(\Mdown, \gdown)$ as long as $\Tup_{ij} \ne 0 $.
From this point of view our proposal for the CCC-scenario using the duality of the conformal scalar fields appears to require some form of fine tuning of initial data away from null infinity to achieve the identification of null infinity and isotropic singularity. 

Suppose we drop the requirement of the coincidence of null infinity and the isotropic singularity in the bridging spacetime $(M,g)$. In this case the vanishing of $\phiup$, on some surface $S$ say, still triggers the switch to the dual spacetime, which once more contains an isotropic singularity (Lemma \ref{Lemma isotropic singularity}). Note that in $(M,g)$ the past aeon and the next aeon coexist for some conformal time until the past aeon has reached null infinity. It is unclear how close $S$ should be to null infinity (with respect to the bridging metric $g$) to assure that all black holes have already evaporated\footnote{Note that \cite{Ellis no pop} argues that black holes will not pop (completely evaporate).}? Or what would happen in $(\Mdown, \gdown)$ if $S$ contained trapped surfaces in $(\Mup, \gup)$? Recall that the conformal scalar field may violate the dominant energy condition, so that singularity theorems may not be applied. We will not pursue these kind of questions here.

The focus of this article has been to identify an explicit mathematical formulation that can describe how the new aeon in the CCC-scenario can be generated from the current aeon. In our proposal the conformal factor relating these two aeons is determined by the physical conformal field $\phiup$. It is thus independent of the choice of bridging metric or the reciprocal hypothesis. 

Indepentently of the discussion of the CCC-scenario, we have extended the work of \cite{Bekenstein1974} to include cosmological constants and quartic self-interaction terms. As a result we observed that under Bekenstein's duality the cosmological constant and the coefficient of the quartic switch roles.
We have shown the existence, uniqueness and stability of weakly asymptotically flat spacetimes containing a conformal scalar field, Yang-Mills fields and radiation fluids (Theorem \ref{Existence Theorem}). Moreover, we have proven the existence of a large class of spacetimes with the above combination of matter models that contain an isotropic singularity modelling the big bang. As highlighted above the matter contents can be generalised to other trace-free matter models if suitable formulations of the CEFE can be obtained.

\section*{Acknowledgements}
The author would like to thank UCL for Visiting Research Fellowship and JSPS for a Research Fellowship. Further, the author would like to acknowledge financial support by the grant CERN/FP/123609/2011 of FCT and CERN.
The author has benefitted from discussions with J.A. Valiente Kroon and K.P. Tod.


\end{document}